\documentclass[man,floatsintext]{apa7}
\usepackage[american]{babel}
\usepackage{csquotes}
\usepackage{amsmath, amssymb, mathtools}
\usepackage{graphicx}
\usepackage{etoolbox}
\usepackage{booktabs}
\usepackage{caption}
\usepackage{siunitx}
\usepackage{xfrac}
\usepackage{listings}
\usepackage{xcolor}
\usepackage{newtxtext}
\usepackage{newtxmath}
\usepackage{longtable}\usepackage{booktabs}
\usepackage{longtable}
\usepackage{multirow}
\usepackage{booktabs}
\usepackage{longtable}
\usepackage{multirow}
\usepackage{booktabs}
\usepackage{threeparttable}
\usepackage[style=apa,backend=biber,sortcites=true]{biblatex}
\DeclareLanguageMapping{american}{american-apa}
\AtBeginBibliography{\interlinepenalty=10000}
\title{Modeling Issues with Eye Tracking Data }
\shorttitle{Eye-Tracking Models}
\author{Gregory Camilli}
\authorsaffiliations{Rutgers University}
\authornote {Please send correspondence to greg.camilli@gse.rutgers.edu.}

\abstract{I describe and compare procedures for binary eye-tracking (ET) data. These procedures are applied to both raw and compressed data. The basic GLMM model is a logistic mixed model combined with random effects for persons and items. Additional models address autocorrelation eye-tracking serial observations.  In particular, two novel approaches are illustrated that address serial without the use of an observed lag-1 predictor: a first-order autoregressive model obtained with generalized estimating equations, and a recurrent two-state survival model. Altogether, the results of five different analyses point to unresolved issues in the analysis of eye-tracking data and new directions for analytic development.}

\begin{document}
\maketitle
\keywords{
eye-tracking
\and modeling
\and generalized linear mixed model
\and data compression
\and generalized estimating equations 
\and survival analysis
\and autocorrelation
}

\section{Introduction}Modeling the cognitive processes with which students engage in psychological tasks in eye tracking studies has important implications for theory development. It may also enable the construction of more informative test scores \parencite{DeBoeck2019}. A number of models have been used for the analysis of eye-tracking data \parencite{Barr2008, Chen2019, BrownSchmidt2020}. The basic features of these models include binary or categorical data for eye fixations, and multiple measurements of subject responses over time. The data then manifest as subject-level time series, often composed of Bernoulli observations. \textcite{Cho2018} extended the original work of \textcite{Barr2008} using a generalized linear mixed model (GLMM) to control autocorrelation, a common feature of time series that may bias standard errors. In this paper, I also present an alternative procedure that features \textit{run length encoding} (RLE) combined with a two-state survival model. 

After a brief review of previous studies, two methods of compressing data for eye-tracking analyses are described. In part, compression leads to more efficient analysis and can reduce observed serial correlation in eye-tracking time series. Then I describe six methods for analyzing eye-tracking data, where a method refers to a statistical model combined with a set of raw or compressed data. After analyses are completed, model coefficients are reported with primary attention placed on the experimental effects. Discussion follows concerning further directions for modeling in educational eye-tracking studies.

\section{Relevant Literature}
The original multilevel model for analyzing eye-tracking data was constructed by \textcite{Barr2008} in terms of a cross-classified (subjects by items) design. This logistic regression model included fixed effects for experimental variables. \textcite{Ryskin2015} applied a mixed-effects logistic regression model to cross-classified eye-tracking data to examine how listeners use a speaker’s perspective to identify referents during unscripted conversational instructions. Several studies have updated mixed effects modeling to include effects for autocorrelated data  and time (\cite{cho2018ar, BrownSchmidt2020, ChoTree, BROWNSCHMIDT2025149511,  Cho_Brown-Schmidt_Clough_Duff_2025}). 

The subject observations analyzed in the current study are eye-tracking fixations for visual targets, originally collected by \textcite{Ryskin2015}. The hierarchical structure of the data is characterized by trial-level data nested within subjects and items. For a visual presentation of the data structure, see Figure 3 of \textcite{Cho2018}. On each trial, the speaker and the listener participated together in a brief conversational exchange aimed at identifying a target object (the different objects constitute a set of items). Although they were seated in separate rooms, they communicated naturally through wireless microphones while their eye movements were recorded simultaneously with two remote eye trackers. The variable $Contrast$ captures whether the speaker uniquely identified the target object or whether multiple objects in the listener’s visual scene could match the speaker’s descriptions. It reflects the amount of referential competition the listener must cognitively manage during real-time interpretation. Referential competition was hypothesized to make the task more difficult. The variable $Privileged$ reflects whether listeners could reduce referential competition based on information they infer from the speaker's perspective; this variable was hypothesized to make the task easier.

\textcite{Cho_Brown-Schmidt_Clough_Duff_2025} addressed several issues with GLMM models in eye-tracking studies related to how observations are structured. In particular, binary time series are generated as runs of 0s and 1s, and longer gazes generate longer runs as do higher sampling rates. However, longer runs necessarily result in higher autocorrelation which can affect the quality of estimates and standard errors. For this reason, \textcite{Cho_Brown-Schmidt_Clough_Duff_2025} constructed a generalized mixed model (GLMM) with a lag-1 outcome variable as a predictor in addition to other covariates. This resulted in estimating an autoregression coefficient AR(1) for observed data. Random effects were also included for subject and items as well as the AR(1) coefficient to account for differing levels of autocorrelation by subject and by item. \textcite{Cho_Brown-Schmidt_Clough_Duff_2025} showed that autoregressive GLMM models can recover parameters accurately in a simulation study, but they noted that "alternative methods for modeling AR effects in binary time series data are still in the nascent stages" (p. 652). The goal of this paper is to contribute to this effort.

\section{Data}

The data structure is organized around repeated measures, where eye-tracking observations are collected for subject–item–trial combinations. Each subject completed multiple trials, each associated with a visual display \parencite{Ryskin2015}. Each trial included two perspective-taking conditions ($Contrast$ and $Privileged$) as previously described; these two conditions were assigned as the item level and thus varied across items. In the original data, observations consisted of $n=112$ observations at .01 second intervals. Data are recorded as the number of discrete instances in which a person is gazing at a target area (1) versus off-target (0). The total number of data points usually manifests as strings of 1s and 0s. The original data were presumably recorded at a frequency of 1000Hz and compressed to .01 second intervals. The resulting data file consisted of 995,232 records in long format for 152 subjects.

\subsection{Run Length Encoding}

 Consider an example string of 10 samples of observations taken at equal intervals: 1110001111, where each sample fixation is .01 second long. (Note this is a simplification because recorded data depend to some degree on how data are compressed by the eye-tracking apparatus prior to data file construction.) Assume recorded instances of 1 (on target) and 0 (off target) are collected beginning at start time 0 and ending at 1 second, so that each 0-1 value corresponds to a .1 second time lapse. 

\begin{table}[ht!]
\centering
\caption{Run Length Encoding}
\begin{tabular}{cccc}
\hline
\textbf{Run} & \textbf{y} & \textbf{n} & \textbf{StopTime} \\
\hline
        1 & 1  & 3 & 0.03 sec \\
        2 & 0  & 3 & 0.06 sec \\
        3 &  1  & 4 & 0.10 sec \\
\hline
\end{tabular}
\end{table}

run length encoding (RLE) is illustrated in Table 1. The 10 records are compressed into three, which would seem like a loss of information. However, knowing the bin size is .01 seconds, or the total time interval and number of bins, would allow the original data set to be reconstructed in its entirety. This suggests there is no more or less information in the compressed than in the original data. Thus, RLE is lossless compression. The RLE data are used with the survival model described below.

\subsection{Data Selection for Analysis}
 I modified the data by excluding two types of subject response. First, in the original study there were multiple trials in which the same item could be presented more than once for a subject. I retained only the first trial of an item for each subject. Second, there were subject-by-item time series with only a single gaze recorded. In many of these instances, the subject appeared to be uniformly gazing off-target for two 2-3 items in a row. This type of data is commonly considered "missing" in eye-tracking studies, which is related to long off-screen periods \parencite{rayner2009eyemovements, staub2013individual}. These particular series were deleted from the data  in the current study. This selection procedure resulted in an uncompressed long-format data set of 646,016 records. An RLE data set was then constructed as a long-format data set of 15,025 records (a compression rate of about 98\%). 

\section{Modeling}
Modeling procedures in this study are applied to the data collected by \textcite{Ryskin2015} and previously analyzed by Cho et al. (2018). The data described above were analyzed by \textcite{Cho2018} using a generalized linear mixed model. This model included a lag-1 observed outcome as a predictor to account for autocorrelation. Other predictors included experimental variables. The model also incorporated random effects for subjects and items. In this paper, I use several similar models for data obtained with the exclusions described previously. I also added the two interaction terms $Time*Privileged$ and $Time*t$. These interactions are designed to detect whether the effects of the experimental variables change across time. 

The first model was run on the full uncompressed data set. I refer to this as the GLM model, which is given by:

\begin{equation}
\begin{aligned}
Y_{ijt} &\sim \text{Bernoulli}\!\left(p_{ijt}\right), \\[1mm]
\text{logit}\!\left(Y_{ijt} = 1 |p_{ijt}\right) 
  &= \mu
   + \beta_1\,\text{Contrast}_{ij}
   + \beta_2\,\text{Privileged}_{ij} \\
  &\quad
   + \beta_3\,\text{Time}_{t}*\text{Contrast}_{ij}
   + \beta_4\,\text{Time}_{t} *\text{Privileged}_{ij} \\
  &\quad
   + u_i + v_j , \text{where} \\
u_i &\sim N(0, \sigma_u^2), \\
v_j &\sim N(0, \sigma_v^2).
\end{aligned}
\end{equation}

\noindent Here, the binary response $y_{ijt}$ for subject $i$ and item $j$  at time $t$ is modeled using a Bernoulli regression with fixed effects for t, Privileged, the interaction of $Time$ with $Contrast$ and $Privileged$, and random effects for subject and item.  The LAG model was obtained by augmenting the GLM model with a lag-1 predictor $y_{ij,t-1}$ to account for serial recursion:

\begin{equation}
\begin{aligned}
 Y_{ijt} &\sim \text{Bernoulli}\!\left(p_{ijt}\right), \\[1mm]
\text{logit}\!\left( Y_{ijt} = 1 | p_{ijt}\right) 
  &= \mu
   + \beta_0\,\text{y}_{ij,t-1}
   + \beta_1\,\text{Contrast}_{ij}
   + \beta_2\,\text{Privileged}_{ij} \\
  &\quad
   + \beta_3\,\text{Time}_{t}\!*\!\text{Contrast}_{ij}
   + \beta_4\,\text{Time}_{t}\!*\!\text{Privileged}_{ij} \\
  &\quad
   + u_i + v_j ,
\end{aligned}
\end{equation}

\noindent where the random effects $u$ and $v$ share the same variance structure in Equation 2.

The GLM model is altered more substantially by omitting random effects and incorporating a coefficient $\phi_A$ as an alternative for the lag-1 autocorrelation. Error correlation is then incorporated directly into the variance structure. To show the application of this approach, generalized estimating equations can be  used. To show the application of this structure, it is expanded in some detail.

\section{Generalized Estimating Equations.}
Generalized Estimating Equations (GEE) \parencite{LiangZeger1986}
provide a semiparametric approach for marginal regression with correlated errors. For binary outcomes, let $p_{ijt} = \Pr(Y_{ijt} = 1)$ denote the marginal mean for subject $i$ for item $j$ at time $t$. The mean model is written as

\begin{equation}
\begin{aligned}
Y_{ijt} &\sim \text{Bernoulli}\!\left(p_{ijt}\right), \\[1mm]
\text{logit}\!\left(Y_{ijt} = 1 | p_{ijt}\right) 
  &= \mu
   + \beta_1\,\text{Contrast}_{ij}
   + \beta_2\,\text{Privileged}_{ij} \\
  &\quad
   + \beta_3\,\text{Time}_{t}\ * \text{Contrast}_{ij}
   + \beta_4\,\text{Time}_{t}\ * \text{Privileged}_{ij} 
\end{aligned}
\end{equation}

\noindent The marginal variance of each Bernoulli observation is
$\sigma_{ijt}^2 = \sigma^2(Y_{ijt}) = p_{ijt}(1 - p_{ijt})$, and the variance-function matrix for cluster $ij$ is
$A_{ij} = \mathrm{diag}(\sigma_{ij1}^2, \ldots, \sigma_{ijT}^2)$, where $T=112$. 
Serial or within-cluster correlation is introduced by specifying a working correlation matrix $R$ containing the single parameter $\phi_A$. For AR(1) structure, $R$ is a symmetric matrix with off-diagonal elements defined as $\phi_A^{|l-m|}$, where $l$ and $m$ are two time indices. The observed vector $Y_i = (Y_{ij1},\ldots,Y_{ijT})^\top$ then has the variance structure:

\begin{equation}
\begin{aligned}
V_{ij} &= \alpha A_{ij}^{1/2}\, R \, A_{ij}^{1/2},
\end{aligned}
\end{equation}

\noindent where $\alpha$ is the scale parameter. Thus, the residual correlation $\phi$ enters the estimating equations through
$R$.

Let
\[
D_{ij} = \frac{\partial p_{ij}}{\partial \beta^\top}
\]
be the $T \times q$ matrix of derivatives of the mean vector (where $q = 6$) with
respect to $\beta$.  The GEE estimator $\hat{\beta}$ is then obtained by solving the quasi-score equation

\begin{equation}
\sum_{i=1} 
D_{ij}^{\top} V_{ij}^{-1} (Y_{ij} - p_{ij}) = 0.
\end{equation}

Estimation of the working correlation parameter $\phi$ is performed by
a moment equation based on Pearson residuals,
\[
r_{it} = \frac{Y_{ijt} - p_{ijt}}{\sigma_{ijt}},
\qquad
r_{ij} = (r_{ij1},\ldots,r_{ijT})^\top.
\]
For an $\mathrm{AR}(1)$ structure, a common estimator is

\begin{equation}
\hat{\phi}
 = 
 \frac{\displaystyle \sum_{ij} \sum_{t=2}^{T} r_{ijt}\, r_{{ij},t-1}}
      {\displaystyle \sum_{ij} (T - 1)},
\end{equation}

\noindent where $\hat{\phi_A}$ is the residual correlation, and the large-sample covariance of $\widehat{\beta}$ is estimated using the robust sandwich estimator is

\begin{equation}
\begin{aligned}
\hat{\sigma}^2(\hat{\beta})
  &= B^{-1} M B^{-1}, \text{where}\\[6pt]
B &= \sum_{ij} D_{ij}^{\top} V_{ij}^{-1} D_{ij}, \\[6pt]
M &= \sum_{ij} 
      D_{ij}^{\top} V_{ij}^{-1}
      (Y_{ij} - \hat{p}_{ij})(Y_{ij} - \hat{p}_{ij})^{\top}
      V_{ij}^{-1} D_{ij} .
\end{aligned}
\end{equation}

\noindent In Equations (5) and (7) it can be seen that $\phi_A$ affects both fixed-effect estimates and standard errors.

Several authors have noted that the GEE working covariance $V_i = \alpha A_i^{1/2} R_\phi A_i^{1/2}$ plays a role analogous to the autocorrelation of latent residual processes in GLM or state-space models for binary data.  In particular, \parencite{ZegerLiangAlbert1988} demonstrated that serial correlation among binary observations can be represented either by specifying a latent Gaussian process with AR(1) correlation or by imposing the same AR(1) structure on the working correlation matrix in GEE. Related comparisons appear in \parencite{HeagertyZeger2000}, who show that marginal models in GEE and latent-process conditional models often share identical first- and second-moment structures, though they differ in higher-order behavior and interpretation.

Rather than controlling for autocorrelation with an observed lag-1 indicator, $\phi_A$ can me incorporated directly into the model's variance structure. Residual structures are more likely to represent a stationary process, especially when $Time$ is included as a covariate. \textcite{Levin2018ChangeBlindness} previously applied a GEE model to analyze eye-tracking data; however, they used an unstructured correlation matrix as opposed to a first-order error structure. The AR1 variance structure achieves a higher degree of efficiency by estimating a single coefficient $\phi_A$, but this economy depends on the length of the time series.  

\section{Survival Modeling}
RLE modeling of eye tracking responses could be used to model the 0-1 category frequencies, rather than the 0-1 data individually. This strategy was suggested previously in the discussion of Table 1. While binomial modeling can be used in the framework of a GLMM. However, it this results in a serial correlation of $\rho = -1$. It is more interesting to treat the runs treated as discrete gaze episodes that occur over time. For RLE compression, the blocks correspond to $Y=0$ or to $Y=1$. After compression, each episode can be represented by an interval $[Time_{\text{start}}, Time_{\text{stop}}]$ and an associated \textit{event} indicator that marks whether a transition to the opposite gaze state occurred at the end of the interval. This representation lends itself to an alternating two–state survival framework, where separate transition hazards are modeled for leaving the target $1 \rightarrow 0$ and entering the target $0 \rightarrow 1$  \parencite{kleinbaum2012survival, therneau2024survival}. Because RLE compression preserves the dwell structure, the resulting episode durations provide suitable survival intervals for Cox regression. The COX model is:

\begin{equation}
\begin{aligned}
h(t \mid X, S)
  &= h_{0,S}(t)\,
     \exp\Bigl(
        \beta_1\,\text{Privileged}
        + \beta_2\,\text{Contrast} \\
  &\quad
        + \beta_4\,\{ 0\!\to\!1\} * \text{Privileged} \\
  &\quad
        + \beta_5\,\{0\!\to\!1\} * \text{Contrast}
     \Bigr)
\end{aligned}
\end{equation}

\noindent In a recurrent two--state process with states $\{0,1\}$, the hazard function $h_{S \to S'}(t)$ describes the instantaneous rate of transition from state $S$ to state $S'$ at time $t$, conditional on the process having remained in state $S$ up to time $t$. Equivalently, the hazard quantifies how quickly a transition is expected to occur at time $t$, given that the system is currently in state $S$.

In the fitted model, the baseline hazard is stratified by transition type so that the temporal risk profiles for leaving and entering the target object may differ freely. Predictors such as \textit{Privileged} and \textit{t} are allowed to influence each transition differently through interaction terms with the transition indicator. This structure has interpretable effects: the main coefficients describe how covariates affect the hazard of leaving the target, while the interaction coefficients quantify how those effects change during target reacquisition. For this analysis, the main focus is on the effect for the $0 \rightarrow 1$ transition. The survival model with RLE-encoded data provides a computationally efficient modeling alternative to AR1 with uncompressed data. 

The model incorporates a Cox proportional hazards framework, the process being modeled is not survival in the conventional sense. The data consist of a binary time series in which each subject–item trajectory alternates repeatedly between two states, with each interval representing the duration spent in a given state before switching to the other. In this context, the COX model is used as a flexible estimator of state-specific switching intensities rather than time to a terminal event. By stratifying on transition direction, separate baseline hazards are used for $0 \rightarrow 1$ and $1 \rightarrow 0$ transitions, so that covariate effects are interpreted as changes in the instantaneous rate of leaving the current state.

A related model to the previous one is an ON–OFF survival model, which treats a binary time series as a sequence of state episodes (runs) and models the time spent in each state as a survival outcome. Each run begins when the process enters the ON or OFF state and ends when it switches to the other state. The survival time is the duration in the current state, and the hazard represents the instantaneous risk of switching given how long the state has lasted. By stratifying on transition type (ON→OFF vs OFF→ON), the model allows different baseline hazards for the two states. Covariates can modify these hazards to lengthen or shorten ON or OFF durations. When the time axis is reset at each state entry, the model is semi-Markov, focusing on duration dependence rather than absolute trial time; absolute time can be added as a covariate if needed to capture learning or fatigue. Experimental coefficients describe how the manipulation lengthens or shortens the duration of ON (or OFF) states, not how often states occur or when in the trial switches happen.

\section {Results}

Analyses of the GLM and LAG methods were carried out with the R function glmer from the package lme4 \parencite{Bates2015_lme4}. The R function glmgee from the R package glmtoolbox \parencite{Vanegas2023} was used for analysis of the AR1 model, and the R function coxph from the package survival was used for the two-state COX model \parencite{therneau2024survival}. The AR1 model produced positive definite $R$ matrices; however, the determinants were near zero. To address this issue, user-defined $R$ matrices were constructed with $\phi = .95$ with a ridge of 1e-5. For AR1, estimates were nearly identical to those produced in runs when $\phi$ was freely estimated. 

Because of the data exclusion procedures used in the present study, analytic results cannot be directly compared to prior studies. The data for the analysis were modified in two ways. First, models using a lag-1 predictor required dropping the first observation in each individual time series. Second, for the survival models, the last run was deleted because the transition indicator was missing by fiat. To account for non-independence across subject and items, and a subjects-by-item indicator was used as a cluster variable in the AR1, COX, and ON-OFF models. 

\begin{table}[ht]
\caption{Fixed-effect estimates for five methods.}
\centering
\begin{tabular}{lccccc}
\hline
Term & GLM & AR1 & LAG & COX & ON-OFF\\
\hline
Intercept    & -0.437 & -0.401 & -2.946 &    --    &   ---   \\
Ylag-1       &  --    &  --    &  5.784 &    --    &   ---   \\
Privileged   &  0.076 &  0.071 &  0.083 &    0.070 &   0.069 \\
Contrast     & -0.079 & -0.068 & -0.013 &   -0.059 &  -0.077 \\
Time         &  0.854 &  0.831 &  0.498 &    --    &   1.816 \\
Priv*Time    & -0.012 &  0.006 &  0.012 &    --    &   ---   \\
Contrast*Time& -0.015 & -0.012 &  0.028 &    --    &   ---   \\
\hline
\end{tabular}
\begin{tablenotes}[flushleft]
\small
\item \textit{Note}.The 0-to-1 transition effects are reported for the survival models.
\end{tablenotes}
\end{table}

\subsection{Fixed Effects} The estimates of fixed effects for these models are given in Table 2. For the survival analyses, the transition interaction ($ y=0 \rightarrow y=1)$ effects are given. The LAG model incorporating the lag-1 $y$ predictor stands out against the other four models. In particular, the LAG coefficients for $Time$ and $Contrast$ are smaller, and the coefficient for $Privileged$ is higher. A $Contrast*Time$ interaction is also evident for the lag model. The GLM, AR1, and ON-OFF models indicate a strong positive relationship between time and the likelihood of being on-target, but the lag-1 $Time$ coefficient is an order of magnitude higher. The variable $Time$ is implicitly contained in the COX model through run start and stop times. For this reason, the COX method did not produce intercepts or time interaction.

It is interesting that the GLM, AR1 and COX/ON-OFF models resulted in similar estimates  of experimental effects as the GLM model. Whereas the GLM and AR1 procedures estimate the probability of a on-target gaze, the survival models estimate the transition from on off-target gaze to an on-target gaze. 

\subsection{Standard Errors} Standard errors for the fixed effects are given in Table 3. The models provide notably dissimilar SEs. The GLM and LAG methods provided the lowest, and the AR1 and survival SEs were about double of those for the GLM model. However, the survival models had lower effective sample size because, as noted asbove, the last run in each individual time series is dropped due to the lack of a transition indicator.  

\begin{table}[ht]
\caption{Standard errors of fixed-effect estimates for five methods.}
\centering
\begin{tabular}{lccccc}
\hline
Term & GLM & AR1 & LAG & COX & ON-OFF\\
\hline
Intercept     & 0.035 & 0.016 &  0.039   & ---   & ---   \\
Ylag-1        & ---    & ---  &  0.012   & ---   & ---   \\
Privileged    & 0.007 & 0.037 &  0.014   & 0.050 & 0.053 \\
Contrast      & 0.004 & 0.023 &  0.009   & 0.029 & 0.031 \\
Time          & 0.003 & 0.015 &  0.006   & ---   & 0.053 \\
Priv*Time     & 0.007 & 0.035 &  0.015   & ---   & ---   \\
Contrast*Time & 0.005 & 0.022 &  0.009   & ---   & ---   \\
\hline
\end{tabular}
\begin{tablenotes}[flushleft]
\small
\item \textit{Note}. Robust SEs for survival models are reported for the 0-to-1 transition for the experimental effects.
\end{tablenotes}
\end{table}

\subsection{Random Effects, $\phi$, and Concordance}
Random effects are given in Table 4 for the GLM and LAG models. For the AR1, COX, and ON-OFF models, the cluster option was used in lieu of adding random effects. The estimated variance components obtained for the GLM and LAG methods were moderately similar. For the AR1, the estimated residual correlation was extremely high. However, this is not unexpected given the median run length of 35 (out of 112) across all subjects. This level of $\phi$ raises the concern that other coefficients may be destabilized. This is also true for the LAG model because a large pseudo $R^2$ is implied by the high magnitude of the $Ylag1$ coefficient. The concordance coefficient was higher for the ON-OFF model (.574 v. .512), meaning the latter model is noticeably better at ranking runs by time-to-switch. Thus, runs that switched earlier were assigned noticeably higher predicted risk (transition to 1) with the ON–OFF model.

\begin{table}[ht]
\caption{Random-effects variance, concordance, and associated parameters.}
\centering
\begin{threeparttable}
\begin{tabular}{lccc}
\toprule
Model & Subj Var & Item Var & Concordance\\
\midrule
GLM   & 0.071 & 0.069  & ---  \\
LAG   & 0.107 & 0.074  & ---  \\
COX   & ---   & ---    & 0.512\\
ON-OFF& ---   & ---    & 0.574 \\
\bottomrule
\end{tabular}
\end{threeparttable}
\begin{tablenotes}[flushleft]
\small
\item \textit{Note}. Ridge estimates are reported for AR1 with $\phi = .95$. Estimated value of $\phi = .963$ and estimated $\alpha = 1.001$. Concordance SEs are both .003.
\end{tablenotes}
\end{table}

\section{Discussion}

The results of this study may shed some light on the issue of autocorrelation in eye-tracking data. Relative to the GLM model, the AR1 and survival models provided more about the same estimates of experimental effects. This result raises important and unresolved questions. First, the finding that AR1 and COX/OFF-OFF models provided similar estimates of experimental effects may not be a coincidence. Recent work has shown that binary time–series models with autoregressive structure can behave very much like discrete-time two–state transition or survival model \parencite{campajola2020ising, gorgi2016survival}. It is an critical caveat that the AR1 model is not truly time series models but rather an error-correlation model. Second, while the autoregressive model with lag coefficients is a useful exploratory tool, it may be impractical for longer time series due to increasing multicollinearity. While a upper boundary can be set manually for the autocorrelation and a ridge added to $|R|$, these are purely ad hoc solutions. The survival models, or similar models, may provide a way to avoid near singularities due to serial correlation. The survival models, with their relationship to Markov models, seem more interesting for further research. 

A third issue is that for hierarchical time-series models, it is well documented that autoregressive parameters and random effects are difficult to identify simultaneously. The coefficients $\phi$ of AR1 and $\beta_0$ of the LAG model can account for the same temporal structure as the subject and item random intercepts. When the available information is insufficient to distinguish their contributions, the model may distribute this dependence ambiguously between them, producing weak identification of $\phi$ and $\beta_0$, and unstable or inflated variance estimates for the random effects \parencite{diggle2002, galecki2013,}. These difficulties arise particularly when individual time series are short \parencite{verbeke2000, shumway2017, jones1993}. For these reasons, methodological guidance commonly recommends avoiding the joint modeling of autocorrelation and random effects unless the study design provides strong and independent information to identify both components \parencite{pinheiro2000}. It is possible that a similar problem arises from modeling with a lag-1 dependent variable. For instance, in Table 2 the lag-1 coefficient may have redistributed $Time$ and $Contrast$ effects to the $Contrast*Time$ interaction (and subject variance component as shown in Table 4). 

A fourth issue is the key question of what outcome to model. The GLM, AR1, and LAG models estimate the probability of an on-target gaze, while the survival models focus on the transition from off-target to on-target gazes. This would appear to be a substantive question, and it has important consequences for obtaining standard errors. It would seem less important to model probabilities for adjacent gazes in a run rather than to discover which factors facilitate the transition to an on-target gaze. In this regard, the LAG model initially bears some resemblance to a logistic Markov model, but the resemblance is superficial because a lag-1 indicator does not represent a state transition when binary data are composed of runs. Finally, the strong similarity between the GLM and Survival models for fixed effects may suggest a deeper relationship between modeling on-target gazes and modeling transitions to on-target gazes.

Data pre-processing is another important issue. Eye-tracking systems commonly compress data to avoid creating mammoth output files. Yet even with preliminary machine compression, files may be large. The RLE encoding method for additional compression is convenient for survival analysis and greatly reduces execution time. Pre-processing also includes data cleaning which involves tricky decisions of which experimental records to retain for analysis.
 
The results of this study collectively support the existence of an experimental effect for the variable $Contrast$. For the AR1 and survival models, the effect for $Privileged$ is marginally significant. Also, the interaction terms for GLM and AR1 indicate the effects of $Privileged$ and $Contrast$ are uniform over time. Because estimates of the experimental effects in the current study varied in statistical strength across modeling approaches, this demonstrates model assumptions can have substantively important consequences. 

The current study is based on a single data set, yet it does raise conceptual issues in the statistical modeling of eye-tracking data. Results are limited to a two-regime model for on-target and off-target, and a single data set. However, useful directions are suggested for future research on modeling issues in eye-tracking data. In particular, Markov transition models are suggested, but suitable software functions for the current data set would require capability for incomplete subject-by-item designs as well as clustering. 

\section{Acknowledgments}
I used AI tools in the preparation of this paper. This included preparation of references in biblatex format, review of relevant psychological literature for understanding experimental variables, querying regarding modeling issues, suggesting R code for data file preparation, and obtaining latex code for  tables and equations. I generated three paragraphs of text which were heavily edited prior to finalizing the paper. I also used AI to revise a handful of individual sentences.

\printbibliography

@article{cho2018ar,
  title        = {Autoregressive generalized linear mixed effect models with crossed random effects: An application to intensive binary time series eye-tracking data},
  author       = {Cho, Sun{-}Joo and Brown-Schmidt, Sarah and Lee, Woo-Yeol},
  journal      = {Psychometrika},
  volume       = {83},
  number       = {3},
  pages        = {751--771},
  year         = {2018},
  publisher    = {Springer},
  doi          = {10.1007/s11336-018-9604-2},
  url          = {https://doi.org/10.1007/s11336-018-9604-2}
}

@article{Barr2008,
  author       = {Barr, D. J.},
  title        = {Analyzing 'visual world' eyetracking data using multilevel logistic regression},
  journal      = {Journal of Memory and Language},
  year         = {2008},
  volume       = {59},
  number       = {4},
  pages        = {457--474},
  doi          = {10.1016/j.jml.2007.09.002}
}

@incollection{BrownSchmidt2020,
  author       = {Brown-Schmidt, S. and Naveiras, M. and De Boeck, P. and Cho, S.-J.},
  title        = {Statistical modeling of intensive categorical time-series eye-tracking data using dynamic generalized linear mixed effect models with crossed random effects},
  booktitle    = {The Psychology of Learning and Motivation},
  editor       = {Federmeier, K. and Schotter, L.},
  publisher    = {Elsevier},
  year         = {2020},
  volume       = {73},
  doi          = {10.1016/bs.plm.2020.06.004}
}

@article{Chen2019,
  author       = {Chen, Y. and Li, X. and Liu, J. and Ying, Z.},
  title        = {Statistical analysis of complex problem-solving process data: An event history analysis approach},
  journal      = {Frontiers in Psychology},
  year         = {2019},
  volume       = {10},
  pages        = {486},
  doi          = {10.3389/fpsyg.2019.00486}
}

@article{Cho2018,
  author       = {Cho, S.-J. and Brown-Schmidt, S. and Lee, W. Y.},
  title        = {Autoregressive generalized linear mixed effect models with crossed random effects: An application to intensive binary time series eye-tracking data},
  journal      = {Psychometrika},
  year         = {2018},
  volume       = {83},
  pages        = {751--771},
  doi          = {10.1007/s11336-018-9604-2},
  url          = {https://doi.org/10.1007/s11336-018-9604-2}
}

@article{DeBoeck2019,
  author       = {De Boeck, P. and Minjeong, J.},
  title        = {An overview of models for response times and processes in cognitive tests},
  journal      = {Frontiers in Psychology},
  year         = {2019},
  volume       = {10},
  pages        = {102},
  doi          = {10.3389/fpsyg.2019.00102},
  url          = {https://doi.org/10.3389/fpsyg.2019.00102}
}

@article{Ryskin2015,
  author       = {Ryskin, Rachel A. and Benjamin, Aaron S. and Tullis, Jonathan G. and Brown-Schmidt, Sarah},
  title        = {Perspective-Taking in Comprehension, Production, and Memory: An Individual Differences Approach},
  journaltitle = {Journal of Memory and Language},
  year         = {2015},
  volume       = {82},
  pages        = {1--21},
  doi          = {10.1016/j.jml.2015.02.003}
}

@Article{ChoTree,
  author  = {Sun{-}Joo Cho and Sarah Brown{-}Schmidt and Paul De Boeck and Jianhong Shen},
  title   = {Modeling Intensive Polytomous Time-Series Eye-Tracking Data: A Dynamic Tree-Based Item Response Model},
  journal = {Psychometrika},
  year    = {2020},
  volume  = {85},
  number  = {1},
  pages   = {154--184},
  doi     = {10.1007/s11336-020-09694-6},
}

@article{BROWNSCHMIDT2025149511,
title = {Modeling spatio-temporal patterns in intensive binary time series eye-tracking data using Generalized Additive Mixed Models},
journal = {Brain Research},
volume = {1854},
pages = {149511},
year = {2025},
issn = {0006-8993},
doi = {https://doi.org/10.1016/j.brainres.2025.149511},
author = {Sarah Brown-Schmidt and Sun-Joo Cho and Kimberly M. Fenn and Alison M. Trude}
}

@article{Cho_Brown-Schmidt_Clough_Duff_2025, title={Comparing Functional Trend and Learning among Groups in Intensive Binary Longitudinal Eye-Tracking Data using By-Variable Smooth Functions of GAMM}, 
volume={90}, DOI={10.1007/s11336-024-09986-1}, 
number={2}, 
journal={Psychometrika}, 
author={Cho, Sun-Joo and Brown-Schmidt, Sarah and Clough, Sharice and Duff, Melissa C.}, 
year={2025}, 
pages={628–657}}

@article{rayner2009eyemovements,
  author       = {Rayner, Keith},
  title        = {Eye movements and attention in reading, scene perception, and visual search},
  journal      = {Quarterly Journal of Experimental Psychology},
  year         = {2009},
  volume       = {62},
  number       = {8},
  pages        = {1457--1506},
  doi          = {10.1080/17470210902816461},
}

@article{staub2013individual,
  author       = {Staub, Adrian and Benatar, Ashley},
  title        = {Individual differences in fixation duration distributions in reading},
  journal      = {Psychonomic Bulletin \& Review},
  year         = {2013},
  volume       = {20},
  pages        = {1304--1311},
  doi          = {10.3758/s13423-013-0451-1},
}

@book{diggle2002,
  author    = {Diggle, Peter J. and Heagerty, Patrick J. and Liang, Kung-Yee and Zeger, Scott L.},
  title     = {Analysis of Longitudinal Data},
  edition   = {2},
  year      = {2002},
  publisher = {Oxford University Press},
  address   = {Oxford},
}

@book{verbeke2000,
  author    = {Verbeke, Geert and Molenberghs, Geert},
  title     = {Linear Mixed Models for Longitudinal Data},
  year      = {2000},
  publisher = {Springer},
  address   = {New York},
}

@article{jones1993,
  author    = {Jones, R. H.},
  title     = {Longitudinal Data with Serial Correlation: A State-Space Approach},
  journal   = {Biometrics},
  year      = {1993},
  volume    = {49},
  number    = {4},
  pages     = {1133--1146},
}

@book{pinheiro2000,
  author    = {Pinheiro, Jos{\'e} C. and Bates, Douglas M.},
  title     = {Mixed-Effects Models in S and S-PLUS},
  year      = {2000},
  publisher = {Springer},
  address   = {New York},
}

@book{galecki2013,
  author    = {Galecki, Andrzej and Burzykowski, Tomasz},
  title     = {Linear Mixed-Effects Models Using R},
  year      = {2013},
  publisher = {Springer},
  address   = {New York},
}

@book{shumway2017,
  author    = {Shumway, Robert H. and Stoffer, David S.},
  title     = {Time Series Analysis and Its Applications},
  edition   = {4},
  year      = {2017},
  publisher = {Springer},
  address   = {New York},
}

@article{Bates2015_lme4,
  author  = {Bates, Douglas and M{\"a}chler, Martin and Bolker, Ben and Walker, Steven},
  title   = {Fitting Linear Mixed-Effects Models Using \texttt{lme4}},
  journal = {Journal of Statistical Software},
  year    = {2015},
  volume  = {67},
  number  = {1},
  pages   = {1--48},
  doi     = {10.18637/jss.v067.i01},
}

@manual{therneau2024survival,
  title        = {A Package for Survival Analysis in {R}},
  author       = {Therneau, Terry M},
  year         = {2024},
  note         = {R package version 3.7-0},
  url          = {https://CRAN.R-project.org/package=survival}
}

@book{kleinbaum2012survival,
  author    = {Kleinbaum, David G. and Klein, Mitchel},
  title     = {Survival Analysis: A Self-Learning Text},
  edition   = {3},
  publisher = {Springer},
  year      = {2012}
}

@article{campajola2020ising,
  title        = {On the Equivalence between the Kinetic Ising Model and Discrete Autoregressive Processes},
  author       = {Campajola, Chiara and Lillo, Fabrizio and Mazzarisi, Paolo},
  journal      = {arXiv preprint arXiv:2008.10666},
  year         = {2020}
}

@article{gorgi2016survival,
  title        = {Integer-Valued Autoregressive Models with Survival Probability Driven by a Stochastic Recurrence Equation},
  author       = {Gorgi, Paolo},
  journal      = {arXiv preprint arXiv:1609.01910},
  year         = {2016}
}

@article{LiangZeger1986,
  title={Longitudinal data analysis using generalized linear models},
  author={Liang, Kung-Yee and Zeger, Scott L.},
  journal={Biometrika},
  volume={73},
  number={1},
  pages={13--22},
  year={1986}
}

@article{ZegerLiangAlbert1988,
  title={Models for longitudinal data: a generalized estimating equation approach},
  author={Zeger, Scott L. and Liang, Kung-Yee and Albert, Paul S.},
  journal={Biometrics},
  volume={44},
  pages={1049--1060},
  year={1988}
}

@article{HeagertyZeger2000,
  title={Marginalized multilevel models and likelihood inference},
  author={Heagerty, Patrick J. and Zeger, Scott L.},
  journal={Statistical Science},
  volume={15},
  number={1},
  pages={1--26},
  year={2000}
}

@article{Levin2018ChangeBlindness,
  author       = {Levin, Daniel T. and Seiffert, Adriane E. and Cho, Sun-Joo and Carter, Kelsey E.},
  title        = {Are Failures to Look, to Represent, or to Learn Associated with Change Blindness During Screen-Capture Video Learning?},
  journaltitle = {Cognitive Research: Principles and Implications},
  year         = {2018},
  volume       = {3},
  number       = {1},
  doi          = {10.1186/s41235-018-0142-3}
}

@article{Vanegas2023,
  author  = {Vanegas, Luis Hernando and Rond\'{o}n, Luz Marina and Paula, Gilberto A.},
  title   = {Generalized Estimating Equations using the new R package glmtoolbox},
  journal = {The R Journal},
  year    = {2023},
  volume  = {15},
  number  = {2},
  pages   = {105--133},
  doi     = {10.32614/RJ-2023-056}
}
\end{document}